\documentclass[prd,preprintnumbers,amsmath,amssymb]{revtex4}

\usepackage{bm}
\usepackage{graphicx}

\newcommand{\qmax}{q_{\rm max}}
\newcommand{\radon}[1]{\widehat{#1}}
\newcommand{\uvec}[1]{\widehat{\bf #1}}
\newcommand{\myvec}[1]{{\bf #1}}

\begin{document}

\preprint{CWRU-P10-02}
\preprint{NSF-ITP-02-95}

\title{Recoil momentum spectrum in directional dark matter detectors}

\author{Paolo Gondolo}
\email{pxg26@po.cwru.edu}
\affiliation{Case Western Reserve University, Department of Physics, 10900
  Euclid Ave, Cleveland, OH 44106-7079}

\begin{abstract}
  Directional dark matter detectors will be able to record the recoil momentum
  spectrum of nuclei hit by dark matter WIMPs.  We show that the
  recoil momentum spectrum is the Radon transform of the WIMP velocity
  distribution.  This allows us to obtain analytic expressions for the recoil
  spectra of a variety of velocity distributions. We comment on the possibility
  of inverting the recoil momentum spectrum and obtaining the three-dimensional
  WIMP velocity distribution from data.
\end{abstract}

\maketitle

\section{Introduction}
\label{sec:I}

The identification of dark matter is one of the major open questions in
physics, astrophysics, and cosmology. Recent cosmological observations together
with constraints from primordial nucleosynthesis point to the presence of
non-baryonic dark matter in the universe. The nature of this non-baryonic dark
matter is still unknown.

One of the preferred candidates for non-baryonic dark matter is a weakly
interacting massive particle (WIMP).  Substantial efforts have been dedicated
to WIMP searches in the last decades \cite{review}. A particularly active area
\cite{dm2002} are WIMP direct searches, in which low-background devices are
used to search for the nuclear recoil caused by the elastic scattering of
galactic WIMPs with nuclei in the detector \cite{goodman-witten}. In these
searches, characteristic signatures of a WIMP signal are useful in
discriminating a WIMP signal against background.

A WIMP signature which was pointed out very early \cite{freese} is an annual
modulation of the direct detection rate caused by the periodic variation of the
Earth velocity with respect to the WIMP ``sea'' while the Earth goes around the
Sun.  The typical amplitude of this modulation is 5\%. A modulation with these
characteristics was observed by the DAMA collaboration \cite{dama}, but in light
of recent results \cite{cdms,edelweiss}, its interpretation as a WIMP signal is
currently in question. Different, and possibly clearer, WIMP signatures would
be beneficial.

A stronger modulation, with an amplitude that may reach 100\%, was pointed out
by Spergel in 1988~\cite{s88}. Spergel noticed that because of the Earth motion
around the Sun, the most probable direction of the nuclear recoils changes with
time, describing a full circle in a year.  In particular this produces a strong
forward-backward asymmetry in the angular distribution of nuclear recoils.

Unfortunately it has been very hard to build WIMP detectors sensitive to the
direction of the nuclear recoils. A promising development is the DRIFT detector
\cite{DRIFT}. The DRIFT detector consists of a negative ion time projection
chamber, the gas in the chamber serving both as WIMP target and as 
ionization medium for observing the nuclear recoil tracks. The direction of the
nuclear recoil is obtained from the geometry and timing of the image of the
recoil track on the chamber end-plates.  A 1 m$^3$ prototype has been
successfully tested, and a 10 m$^3$ detector is under consideration.

In addition to merely using directionality for background discrimination, what
can be learned about WIMP properties from the directionality of WIMP detectors?
It is obvious that different WIMP velocity distributions give rise to different
recoil distributions in both energy and recoil direction.  Copi, Heo, and
Krauss \cite{chk}, and then Copi and Krauss \cite{copi-krauss}, have examined
the possibility of distinguishing various WIMP velocity distributions using a
likelihood analysis of the resulting recoil spectra, which they generated
through a Monte Carlo program. They have concluded that a discrimination among
common velocity distributions is possible with a reasonable number of detected
events.

Here we want to gain insight into the properties of the nuclear recoil spectra
in energy and direction. For this purpose, we develop a simple formalism that
relates the WIMP velocity distribution to the distribution of recoil momenta.
We find that the recoil momentum spectrum is the Radon transform of the
velocity distribution (see Eq.~(\ref{eq:main}) below). We apply this analytical
tool to a series of velocity distributions, and discover for example how the
recoil momentum spectrum of a stream of WIMPs differs from that of a Maxwellian
velocity distribution. With our gained insight, we suggest that if a WIMP
signal is observed in directional detectors in the future, it may be possible
to invert the measured recoil momentum spectrum and reconstruct the WIMP
velocity distribution from data.

In Section \ref{sec:II} we describe the general kinematics of elastic
WIMP-nucleus scattering, and in Section \ref{sec:III} we obtain our main
formula for the nuclear recoil momentum spectrum.  Sections \ref{sec:IV} and
\ref{sec:V} contain general considerations and examples of Radon transforms of
velocity distributions. Finally, Section~\ref{sec:inv} discusses the
possibility of inverting the recoil momentum spectrum to recover the WIMP
velocity distribution. The Appendices contain useful mathematical formulas for
the computation and inversion of 3-dimensional Radon transforms.

\section{WIMP-nucleus elastic scattering}
\label{sec:II}

\begin{figure}
  \includegraphics[width=0.8\textwidth]{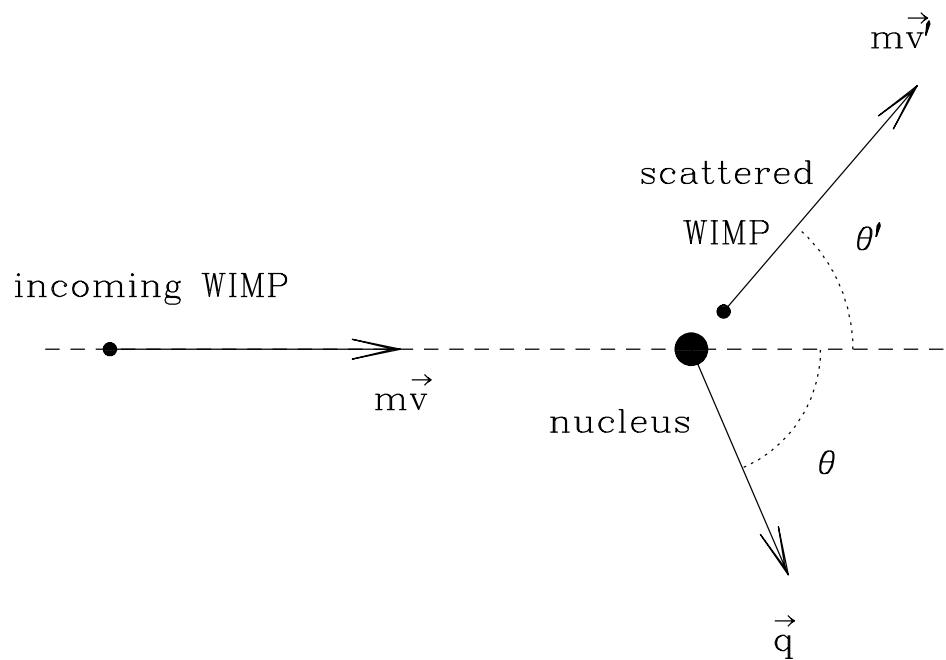}
  \caption{Kinematics of elastic WIMP-nucleus scattering.}
  \label{fig:kinem}
\end{figure}

Consider the elastic collision of a WIMP of mass $m$ with a nucleus of mass $M$
in the detector (see Fig.~\ref{fig:kinem}). Let the arrival velocity of the
WIMP at the detector be $\myvec{v}$, and neglect the initial velocity of the
nucleus.  After the collision, the WIMP is deflected by an angle $\theta'$ to a
velocity $\myvec{v}'$, and the nucleus recoils with momentum $\myvec{q}$ and
energy $E = q^2/2M$.  Let $\theta$ denote the angle between the initial WIMP
velocity $\myvec{v}$ and the direction of the nuclear recoil $\myvec{q}$.
Energy and momentum conservation impose the following relations:
\begin{eqnarray}
&&  \frac{1}{2} m v^2 = \frac{1}{2} m v^{\prime2} + \frac{q^2}{2M} ,
\label{em1} \\
&&  m v' \cos \theta' = m v - q \cos \theta , \phantom{\frac{1}{2}} 
\label{em2} \\
&&  m v' \sin \theta' = q \sin \theta. \phantom{\frac{1}{2}} \label{em3}
\end{eqnarray}
Eliminating $\theta'$ by summing the squares of Eqs.~(\ref{em2}) and
(\ref{em3}),
\begin{equation}
  m^2 v^{\prime 2} = (m v -q\cos\theta)^2 + (q \sin \theta)^2 =
   m^2 v^2 - 2 m v q \cos \theta + q^2 ,
\end{equation}
and using this expression to eliminate $v'$ from Eq.~(\ref{em1}), gives
\begin{equation}
  \label{eq:costheta}
  q = 2 \mu v \cos\theta ,
\end{equation}
where
\begin{equation}
  \mu = \frac{ m M } { m+M } 
\end{equation}
is the reduced WIMP-nucleus mass. We deduce that the magnitude $q$ of the
recoil momentum, and the recoil energy $E$, vary in the range
\begin{equation}
  \label{eq:range}
  0 \le q \le \qmax \equiv 2 \mu v, \qquad 
  0 \le E \le E_{\rm max} \equiv \frac{2 \mu^2 v^2}{M} .
\end{equation}

Eq.~(\ref{eq:costheta}) will be exploited in the following Section to express
the recoil momentum distribution in a simple mathematical form. For this
purpose, we also need the expression for the WIMP-nucleus scattering cross
section. We write the differential WIMP-nucleus scattering cross section as
\begin{equation}
  \label{eq:sigma}
  \frac{d\sigma}{dq^2} = \frac{\sigma_0}{\qmax^2} S(q),
\end{equation}
where $\sigma_0$ is the total scattering cross section of the WIMP with a
(fictitious) point-like nucleus, and $S(q) = | F(q) |^2$ is a nuclear form
factor normalized so that $S(0)=1$. (Both $S(q)$ and $F(q)$ are confusingly
called form factors.) Eq.~(\ref{eq:sigma}) is valid for both spin-dependent and
spin-independent WIMP-nucleus interactions, although $\sigma_0$ and $F(q)$ have
different expressions in the two cases. For example, for spin-independent
interactions with a nucleus with $Z$ protons and $A-Z$ neutrons,
\begin{equation}
  \sigma_0 = \frac{\mu^2}{\pi} \left[ Z G^p_s + (A-Z) G^n_s \right]^2,
\end{equation}
where $G^p_s$ and $G^n_s$ are the scalar four-fermion couplings of the WIMP
with pointlike protons and neutrons, respectively (see Ref.~\cite{lesarcs}).
If the nucleus can be approximated by a
sphere of uniform density, its form factor is
\begin{equation}
  F(q) = \frac{ 9 [ \sin(qR) -q R \cos(qR)]^2 } { (qR)^6 } ,
\end{equation}
where 
\begin{equation}
  R \simeq [0.91 A^{1/3} + 0.3] \times 10^{-13} {\rm cm} 
\end{equation}
is (an approximation to) the nuclear radius. More realistic expressions for
 spin-independent form factors, and formulas for spin-dependent cross sections,
 can be found, e.g., in Refs.~\cite{lesarcs,lewinsmith,jkg,tovey}.

\section{Recoil momentum spectrum}
\label{sec:III}

Eqs.~(\ref{eq:costheta}) and~(\ref{eq:sigma}) can be combined to give the
differential recoil spectrum in both energy and direction, i.e.\ the recoil
{\it momentum} spectrum. We define it as the double differential event rate, in
events per unit time per unit detector mass, differentiated with respect to the
nuclear recoil energy $E$ and the nuclear recoil direction $\uvec{q}$,
\begin{equation}
  \frac{ dR } { dE d\Omega_q} ,
\end{equation}
where $ d \Omega_q$ denotes an infinitesimal solid angle around the direction
$\uvec{q}$.

The double differential rate follows from the double differential cross section
\begin{equation}
  \frac{d\sigma}{dq^2 d\Omega_q} 
\end{equation}
first through the change of differentials $d q^2 =
2M dE$, and then through multiplication by the number $N$ of nuclei in the
detector, division by the detector mass $MN$, and multiplication by the flux of
WIMPs with velocities $\myvec{v}$ in the velocity space element $d^3v$,
\begin{equation}
  n v f(\myvec{v}) \, d^3 v .
\end{equation}
Here $n=\rho/m$ is the WIMP number density, $\rho$ is the WIMP mass density,
and $f(\myvec{v})$ is the WIMP velocity distribution in the frame of the
detector, normalized to unit integral. 

The double differential cross section is obtained as follows. Azimuthal symmetry
of the scattering around the WIMP arrival direction gives $ d\Omega_q = 2 \pi
\, d\!\cos\theta$. The relation between $\cos\theta$ and $q$ in
Eq.~(\ref{eq:costheta}), $\cos\theta=q/2\mu v$, can be imposed through a Dirac
$\delta$ function, $\delta(\cos\theta-q/2\mu v)$. Thus
\begin{equation}
  \frac{d\sigma}{dq^2 d\Omega_q} = \frac{d\sigma}{dq^2} \,\, 
  \frac{1}{2\pi} \,\, \delta\!\left(\cos\theta-\frac{q}{2 \mu v}\right) =
  \frac{\sigma_0 S(q)}{8\pi \mu^2 v} 
  \,\, \delta\!\left(v\cos\theta-\frac{q}{2\mu}\right) .
\end{equation}
This is correctly normalized as can be seen by integration of the expression in
the middle over $d\Omega_q$.  

Summarizing, the double differential event rate per unit time per unit detector
mass is
\begin{equation}
  \frac{ dR } { dE d\Omega_q} = 
  2M \, \frac{N}{MN}  \int
  \frac{d\sigma}{dq^2d\Omega_q} \,\, n \, v \, f(\myvec{v}) \, d^3 v = 
  \frac{n \sigma_0 S(q)}{4\pi \mu^2} \int
  \delta\!\left(v\cos\theta-\frac{q}{2\mu}\right) \, f(\myvec{v}) \, d^3 v .
\end{equation}
We write it as
\begin{equation}
  \label{eq:main}
  \frac{ dR } { dE d\Omega_q} = \frac{ n \sigma_0 S(q) }{ 4 \pi \mu^2 } \, 
  \, \radon{f}(v_q,\uvec{q}) .
\end{equation}
Here
\begin{equation}
  v_q = \frac{q}{2\mu} = \sqrt{ \frac{ME}{2\mu^2} }
\end{equation}
is the minimum velocity a WIMP must have to impart a recoil momentum $q$ to the
nucleus, or equivalently to deposit an energy $E=q^2/2M$, as can be seen from
Eq.~(\ref{eq:range}). Moreover, 
\begin{equation}
  \label{eq:eta}
  \radon{f}(w,\uvec{w}) = \int \delta ( \myvec{v} \cdot \uvec{w} - w) 
  \, f(\myvec{v}) \, d^3 v ,
\end{equation}
is the 3-dimensional Radon transform of the velocity distribution function
$f(\myvec{v})$. We note in passing that $\radon{f}$ has units of inverse speed.

Eq.~(\ref{eq:main}) is the main result of this paper. It states that, apart
from a normalizing factor, the recoil momentum spectrum is the Radon transform
of the WIMP velocity distribution. The Radon transform is a linear integral
transform (see Refs.~\cite{d83,rk96}), which was introduced in two dimensions
by Radon in 1917 \cite{r17}. The Radon transform has been widely studied for
its use in solving differential equations, and especially in two-dimensions,
for its medical applications in computer tomography.  Geometrically,
$\radon{f}(w,\uvec{w})$ is the integral of the function $f(\myvec{v})$ on a
plane orthogonal to the direction $\uvec{w}$ at a distance $w$ from the origin.
For reference, some mathematical properties of the Radon transform are given in
the Appendices.

As a check of our formalism, we show that integrating our basic
equation~(\ref{eq:main}) over recoil directions reproduces the usual expression
for the recoil energy spectrum $dR/dE$.  Applying Eq.~(\ref{eq:etaave}) in 
Appendix A to our expression for the differential rate, we find
\begin{equation}
  \label{eq:usual}
  \frac{ dR } { dE} = \frac{ n \sigma_0 S(q) }{ 2 \mu^2 } \, 
  \, \int_{v>q/2\mu} \frac{f(\myvec{v})}{v} \,\,  d^3v .
\end{equation}
This is the usual expression of the recoil energy spectrum (cfr.\ Eq.~(8.3) in
Ref.~\cite{jkg}).

\section{Computing the recoil momentum spectrum}
\label{sec:IV}

We have cast the nuclear recoil momentum spectrum in terms of a Radon
transform. Now we can take advantage of the properties of Radon transforms,
some of which are listed in the Appendices, to compute recoil momentum spectra
analytically. In this Section we give some general considerations, and in the
next Section we give explicit examples of analytic recoil momentum spectra.

\subsection{Isotropic distributions}

When the WIMP velocity distribution is isotropic, $f(\myvec{v}) = f(v)$, the
recoil spectrum is also isotropic, $\radon{f}(w,\myvec{w}) = \radon{f}(w)$.
From the definition of Radon transform, Eq.~(\ref{eq:eta}),
\begin{equation}
  \label{eq:iso}
  \radon{f}(w) = 2 \pi \int_{w}^{\infty} f(v) v dv .
\end{equation}
We would have obtained the same result starting from Eq.~(\ref{eq:etaave}).

\subsection{Moving observer}

WIMP velocity distributions are often given in the galactic rest frame, while
we are interested in the recoil momentum spectrum in the laboratory frame of
the detector. The change of velocity frame can be performed either on the
velocity distribution before computing the Radon transform or on the Radon
transform computed in the galactic rest frame. The latter is often easier to
compute, and the change of reference frame can be done simply as follows.

The WIMP velocities $\myvec{v}_{\rm lab}$ and $ \myvec{v}_{\rm gal}$ in the
laboratory and galactic rest frames, respectively, are related by
\begin{equation}
  \myvec{v}_{\rm lab} = \myvec{v}_{\rm gal} - \myvec{V}_{\rm lab} ,
\end{equation}
where $ \myvec{V}_{\rm lab} $ is the velocity of the laboratory with respect to
the galactic rest frame. This velocity transformation is a translation in
velocity space, and we can use Eq.~(\ref{eq:transl}) in Appendix A to relate
the Radon transforms in the galactic and laboratory frames,
\begin{equation}
  \label{eq:moving}
  \radon{f}_{\rm lab}(w,\uvec{w}) = 
  \radon{f}_{\rm gal}(w+\myvec{V}_{\rm lab} \cdot  \uvec{w}, \uvec{w}) . 
\end{equation}
Thus the recoil momentum spectrum
in the laboratory frame is given directly in terms of the Radon transform
$\radon{f}_{\rm gal}(w,\uvec{w})$ of the WIMP
velocity distribution in the galactic rest frame by 
\begin{equation}
  \label{eq:etav0}
  \frac{ dR } { dE d\Omega_q} = 
  \frac{ n \sigma_0 S(q) }{ 4 \pi \mu^2 } \, \, \, 
  \radon{f}_{\rm gal}
  \!\left(v_q+\myvec{V}_{\rm lab} \cdot \uvec{q}, \,\uvec{q}\right),
\end{equation}
with $v_q = q/2\mu$ as before.

\subsection{Rotated observer}

If we rotate the coordinate system, we see from Eq.~(\ref{eq:rotate}) in
Appendix A that the recoil momentum spectrum is simply rotated, with the
magnitude of the recoil momentum remaining the same, as expected.

\section{Examples of recoil momentum spectra}
\label{sec:V}

We give some examples of recoil momentum spectra corresponding to common
velocity distributions. We obtain the recoil spectra for streams of particles
and for isotropic and anisotropic Gaussian distributions with and without bulk
velocities.

\subsection{A WIMP stream or flow}

The simplest case is that of a particle stream in which all WIMPs in the stream
move with the same velocity $\myvec{V}$. In this case,
\begin{equation}
  f_{\rm stream}(\myvec{v}) = \delta(\myvec{v} - \myvec{V}), 
\end{equation}
and
\begin{equation}
  \radon{f}_{\rm stream}(w,\uvec{w}) = \delta( \myvec{V} \cdot \uvec{w} - w) .
\end{equation}
The recoil spectrum of a stream with velocity $\myvec{V}$ is concentrated on a
sphere of radius $V/2$, centered in $\myvec{V}/2$ and passing through the
origin. The stream velocity $\myvec{V}$ is a diameter of the sphere. 

\begin{figure}
  \includegraphics[width=0.8\textwidth]{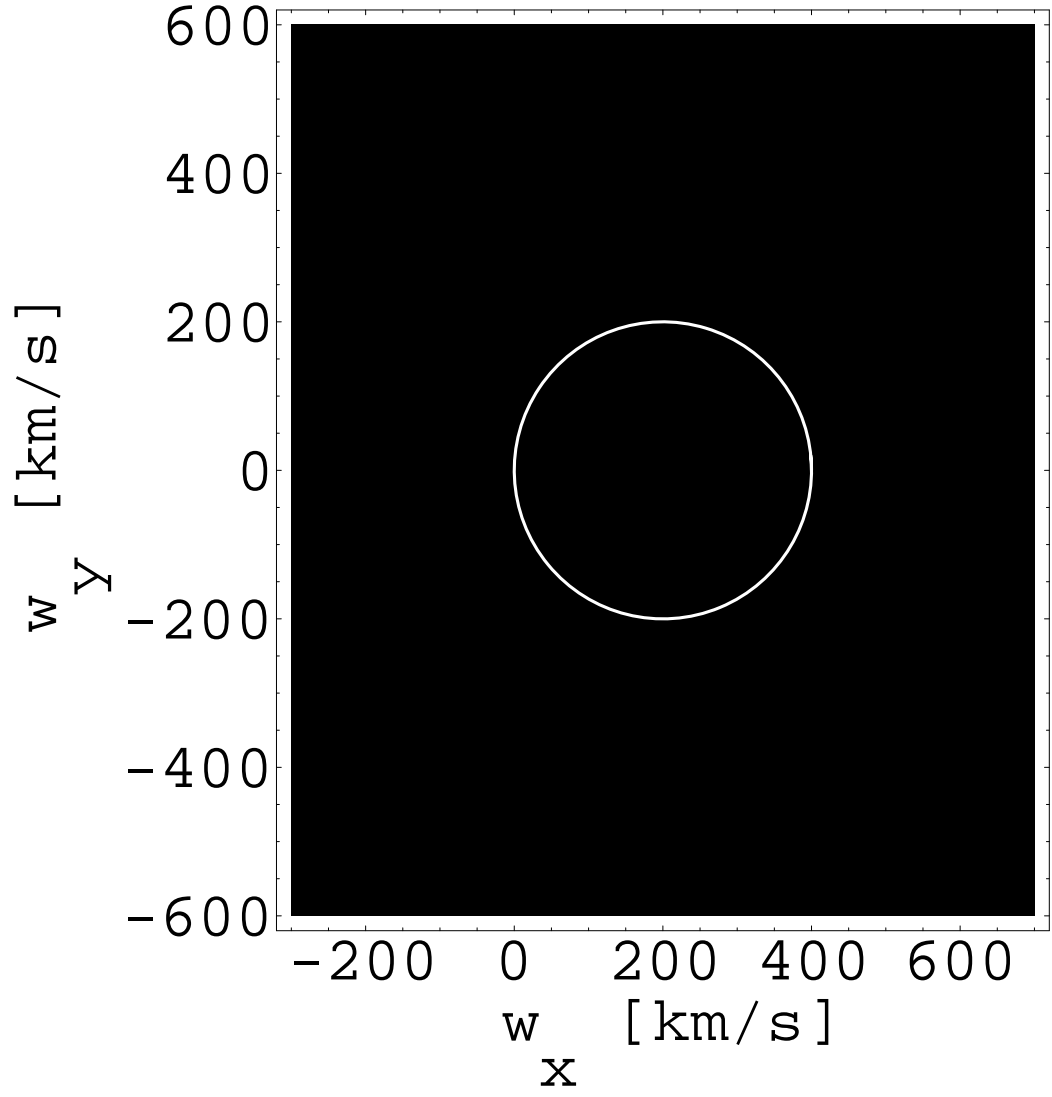}
  \caption{Probability density distribution of the nuclear recoil
    momentum in the recoil plane $(w_x,w_y)$, assuming a stream of WIMPs with
    velocity $(v_x,v_y,v_z) = (400\,{\rm km/s}, 0 ,0)$.  The full
    $(w_x,w_y,w_z)$ distribution can be obtained by revolution around the $w_x$
    axis. The recoil momenta describe a sphere in recoil space.}
  \label{fig:stream}
\end{figure}

Fig.~\ref{fig:stream} shows the $(w_x,w_y)$ section of the recoil momentum
spectrum of a stream of WIMPs arriving from the left with velocity $V_x = 400 $
km/s. The full distribution is obtained through a rotation around the $w_x$
axis. The pattern of recoil momenta forms a sphere.

\subsection{Maxwellian distribution}

A Maxwellian 
distribution with velocity dispersion $\sigma_v$,
\begin{equation}
f_{\rm M}(v) = 
\frac{1}{(2 \pi \sigma_v^2)^{3/2}} 
\exp\left[ {- \frac{v^2}{2\sigma_v^2}} \right], 
\end{equation}
is a particular case of isotropic distribution, and we can use
Eq.~(\ref{eq:iso}) above to compute its Radon transform. We find
\begin{equation}
\radon{f}_{\rm M}(w) = \frac{1}{(2 \pi \sigma_v^2)^{1/2}} 
\exp\left[ {-\frac{w^2}{2\sigma_v^2}} \right] .
\end{equation}

If the detector has velocity $\myvec{V}_{\rm lab}$, we can use
Eq.~(\ref{eq:moving}) to find the Radon transform in the laboratory frame, 
\begin{equation}
  \label{eq:gauss}
  \radon{f}_{\rm M,lab}(w,\uvec{w}) = \frac{1}{(2 \pi \sigma_v^2)^{1/2}} 
  \exp\left[ {-\frac{\left[w+\uvec{w}\cdot\myvec{V}_{\rm lab}\right]^2}
      {2\sigma_v^2}} 
  \right] .
\end{equation}
Notice that $\uvec{w}\cdot\myvec{V}_{\rm lab}$ is the projection of the
velocity of the observer in the direction of the nuclear recoil.  This
expression coincides with, but is simpler than, the analogous expression
obtained by elementary methods in Ref.~\cite{s88} ($\cos\gamma$ in
Ref.~\cite{s88} is $\cos\gamma = - \uvec{w} \cdot \uvec{V}_{\rm lab} $).

\begin{figure}
  \includegraphics[width=0.8\textwidth]{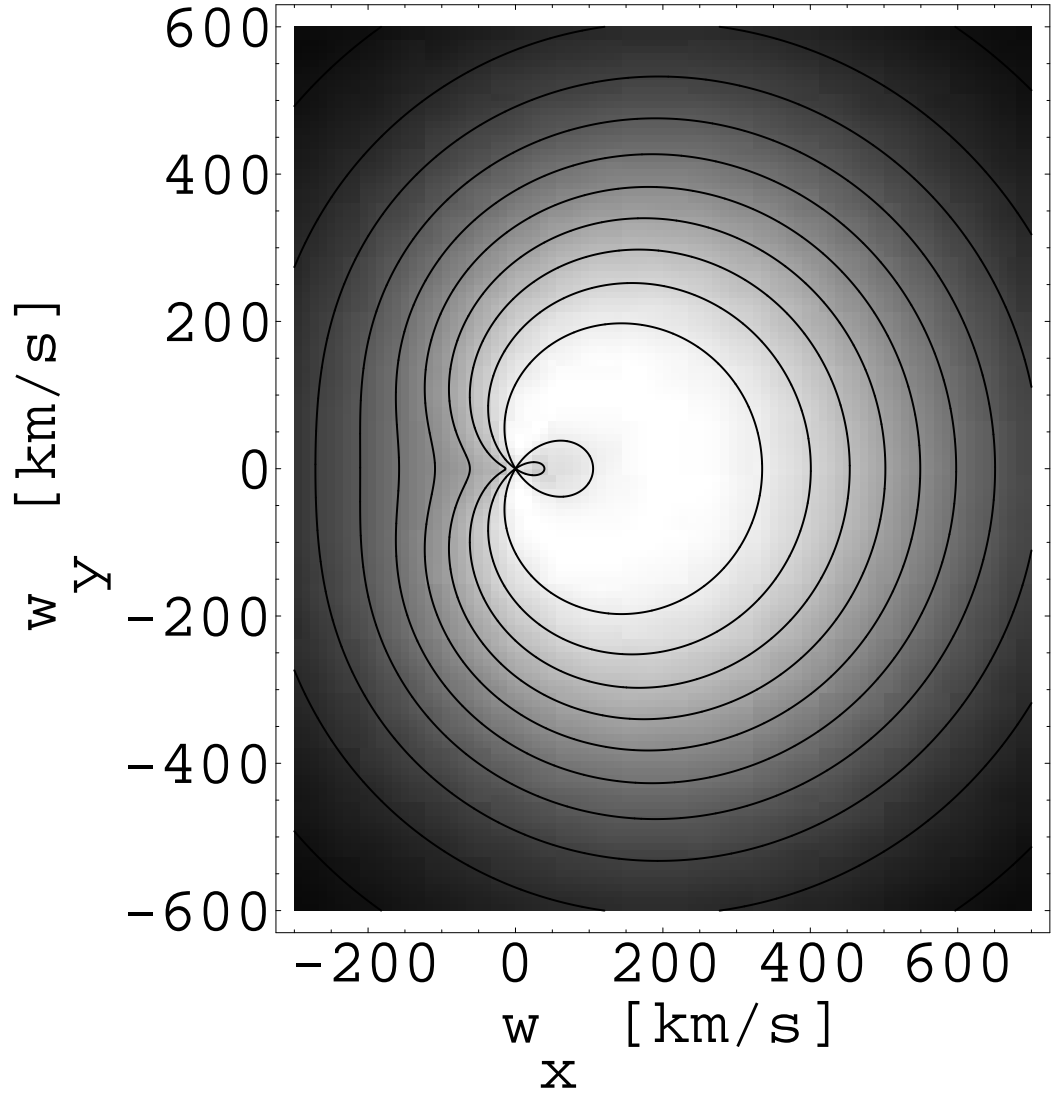}
  \caption{Probability density distribution of the nuclear recoil
    momentum in the recoil plane $(w_x,w_y)$, assuming a Maxwellian velocity
    distribution of WIMPs with velocity dispersion 300 km/s, and a detector
    moving with velocity $(V_x,V_y,V_z) = (-220\,{\rm km/s},0,0)$. Lighter
    areas have higher probability. The full $(w_x,w_y,w_z)$ distribution can be
    obtained by revolution around the $w_x$ axis.}
  \label{fig:maxwell}
\end{figure}

The recoil momentum distribution for a Maxwellian distribution is shown in
Fig.~\ref{fig:maxwell}, assuming a velocity dispersion of 300 km/s and an
observer moving at 220 km/s in direction $-x$. The distribution is symmetric
around the observer velocity. The figure shows the section in the $(w_x,w_y)$
plane only. The full distribution can be obtained by symmetry.

Eq.~(\ref{eq:gauss}) illustrates the reason for writing $\radon{f}(w,\uvec{w})$
instead of $ \radon{f}(\myvec{w})$ (see Section \ref{sec:notation} for more
details). The function $ \radon{f}_{\rm M,lab}(0,\uvec{w}) $ assumes different
values for different directions $\uvec{w}$; the function $ \radon{f}(\myvec{w})
$ would therefore be multivalued at the origin.

\subsection{Truncated Maxwellian distribution}

We may truncate a Maxwellian distribution at the escape speed $v_{\rm esc}$,
\begin{equation}
f_{\rm TM}(v) = \begin{cases}
\displaystyle
\frac{1}{N_{\rm esc} (2 \pi \sigma_v^2)^{3/2}} 
\exp\left[ {- \frac{v^2}{2\sigma_v^2}} \right], & v < v_{\rm esc} \cr
0, & {\rm otherwise} \end{cases},
\end{equation}
with
\begin{equation}
N_{\rm esc} = {\rm erf}\!\left(\frac{v_{\rm esc}}{\sqrt{2}\sigma_v}\right) - 
\sqrt{\frac{2}{\pi}} \frac{v_{\rm esc}}{\sigma_v} 
\exp\left[ {-\frac{v_{\rm esc}^2}{2\sigma_v^2}} \right] .
\end{equation}
Then we have
\begin{equation}
\radon{f}_{\rm TM}(w) = \frac{1}{N_{\rm esc}(2 \pi \sigma_v^2)^{1/2}} 
\left\{ \exp\left[ {-\frac{w^2}{2\sigma_v^2}} \right] -
\exp\left[ {-\frac{v_{\rm esc}^2}{2\sigma_v^2}} \right] \right\}.
\end{equation}

\subsection{Non-isotropic Gaussian distribution}

The recoil-momentum spectrum corresponding to an anisotropic Gaussian
distribution can also be obtained analytically.
An anisotropic Gaussian
distribution with variance matrix $\boldsymbol{\sigma}^{2}$ and 
mean velocity $\myvec{V}$ is given by
\begin{equation}
  f_{\rm gauss}(\myvec{v}) = \frac{1}{(8 \pi^3 \det
    \boldsymbol{\sigma}^2)^{1/2}} 
  \exp\left[ {- \frac{(\myvec{v}-\myvec{V})^T \boldsymbol{\sigma}^{-2} \, 
        (\myvec{v}-\myvec{V})}{2}} \right].
\end{equation}
We are using matrix notation, $\myvec{v}^T$ being the transpose of $\myvec{v}$,
etc.  Using the Fourier slice theorem, actually Eq.~(\ref{eq:fst2}), we find
the Radon transform of the anisotropic Gaussian to be
\begin{equation}
  \radon{f}_{\rm gauss}(w,\uvec{w}) = 
  \frac{1}{(2 \pi \,\uvec{w}^T \boldsymbol{\sigma}^2 \,
    \uvec{w} )^{1/2}} 
  \exp\left[ {-\frac{\left[ w - \uvec{w} \cdot \myvec{V}\right]^2}
      {2\uvec{w}^T \! \boldsymbol{\sigma}^2 \, \uvec{w} }} \right] .
\end{equation}

This is another example of a function which assumes different values at $w=0$
according to the direction $\uvec{w}$.

\section{Reconstructing the velocity distribution}
\label{sec:inv}

The recoil spectrum of a stream and a Maxwellian velocity distribution are very
different: a sphere the first, a smooth distribution the second. This suggests
that it may be possible to distinguish different kinds of WIMP velocity
distributions just by examining the pattern of recoil momenta. Subtle
differences among velocity distributions may be revealed by a maximum
likelihood analysis of the corresponding recoil spectra \cite{chk,copi-krauss}.

More ambitiously, we may think of recovering the WIMP velocity distribution by
inverting the measured recoil momentum spectrum.  Indeed, if we know the
nuclear form factor of the detector nuclei, then for any fixed WIMP mass we can
estimate the Radon transform of the WIMP velocity distribution from the
measured recoil momentum spectrum, modulo a normalization constant $K$.
Eq.~(\ref{eq:main}) can in fact be written as
\begin{equation}
  \radon{f}(v_q,\uvec{q}) = K \, 
  \frac{ 4\pi\mu^2}{S(q)} \, \frac{dR}{dEd\Omega_q},
\end{equation}
enabling us to obtain a measurement of the Radon transform
$\radon{f}(v_q,\uvec{q})$ of the WIMP velocity distribution from the measured
recoil spectrum $ dR/dEd\Omega_q $.  We may be able to invert this Radon
transform and obtain the WIMP velocity distribution $f(\myvec{v})$, again
modulo a normalization constant. Finally, we may be able to fix the
normalization constant either by normalizing $f(\myvec{v})$ to unit integral or
better by examining the detector efficiency as a function of WIMP velocity.

There are several analytic formulas for the inversion of three-dimensional
Radon transforms.  Some of these formulas are collected in Appendix B for
convenience.  Most of the analytical inversion formulas can be converted into
numerical algorithms \footnote{Non-locality complicates the numerical inversion
  of the Radon transform in an even number of dimensions; this difficulty is
  not there in an odd number of dimensions \cite{rk96}.}.  However, any
inversion algorithm we were able to find in the literature is suited only to a
large amount of data in recoil momentum space, since they all assume that it
would be possible to define a discretized version of $\radon{f}(w,\uvec{w})$.
This is {\it not} the case for directional dark matter searches, where the
total number of events is not under the control of the experimentalist and is
expected to be rather small.

New inversion algorithms suited to small numbers of events are therefore needed
if one wants to reconstruct the WIMP velocity distribution using data from
directional detectors. As a first attempt in this direction, we have devised
the following simple algorithm.  Divide the WIMP velocity space into small
cells $S_m$, $m=1,\dots,M$, and assume that the WIMP velocity distribution
$f(\myvec{v})$ is constant over each of these small cells, with value $f_m$ in
cell $S_m$. To each recorded event $j$ with measured recoil momentum
$\myvec{q}_j$, $j=1,\ldots,N$, associate the plane $P_j$ in WIMP velocity space
defined by the equation
\begin{equation}
  P_j : \quad 2 \mu \myvec{v} \cdot \uvec{q}_j = q_j .
\end{equation}
$P_j$ is the plane orthogonal to the recoil direction $\uvec{q}_j$ and at a
distance $w_j=q_j/2\mu$ from the origin. Velocity vectors on this plane are all
the WIMP velocities that can produce the observed nuclear recoil. Let
\begin{equation}
  a_{jm} = {\rm area}( S_m \cap P_j ) ,
\end{equation}
the area of the intersection of the plane $P_j$ with the cell $S_m$ (see
Appendix C for an explicit expression). For each event $j$, assign weight
$a_{jm}$ to the $m$-th cell. Sum the weights over the events, $A_m =
\sum_{j=1}^{N} a_{jm}$, essentially counting how many planes cross any given
cell. Take the discrete Laplacian of the sum of the weights, and keep only
those cells whose values exceed a predetermined threshold.  The resulting
distribution of cell values is our estimate of the WIMP velocity distribution.

\begin{figure}
  \includegraphics[width=0.8\textwidth]{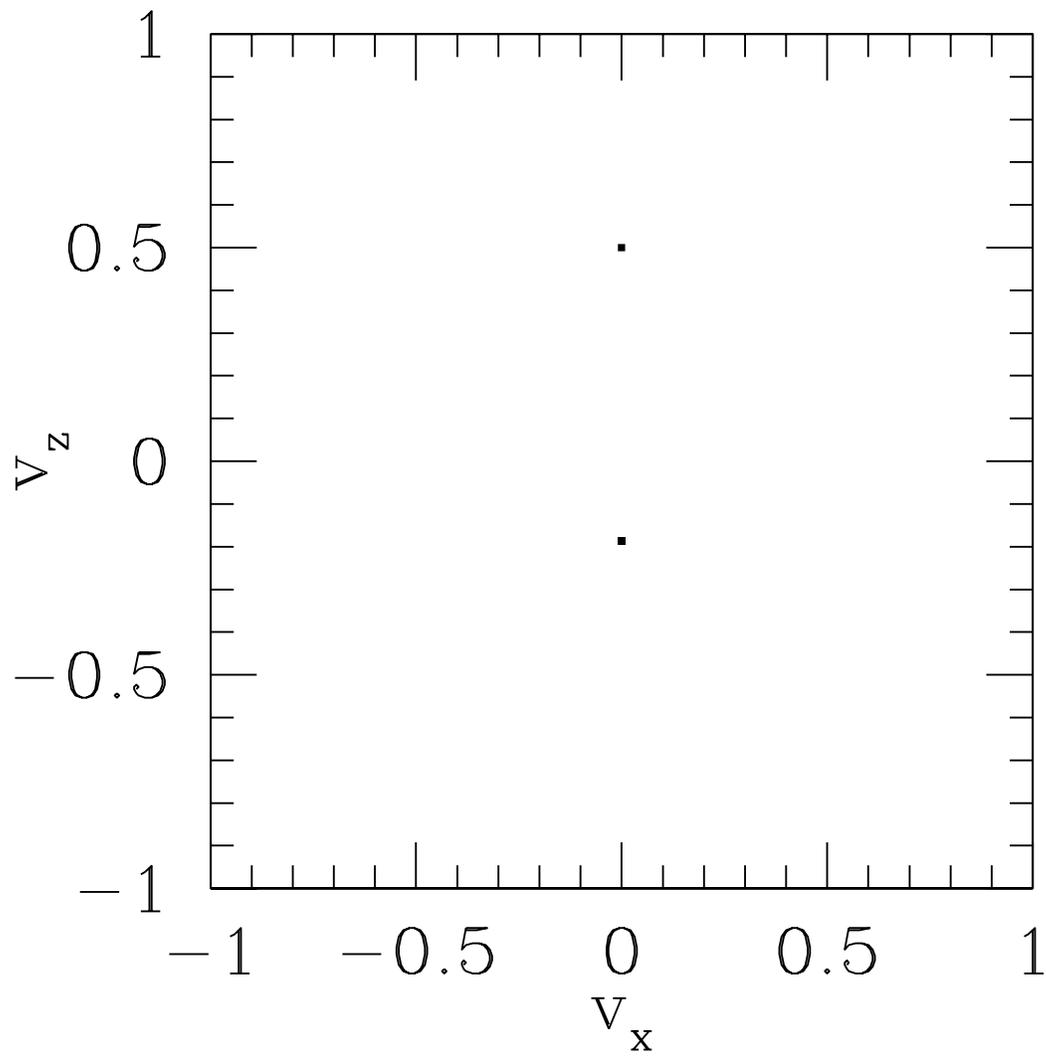}
  \caption{Reconstructed velocity distribution of two WIMP streams with
    velocities $(0,0,0.5)$ and $(0,0,-0.2)$ (in arbitrary units). Only the
    $(v_x,v_z)$ section is shown.}
  \label{fig:pignose}
\end{figure}

To test the capabilities of our algorithm, we simulated the recoil spectrum due
to two streams of WIMPs arriving at the detector from opposite directions, with
velocities $(V_{1x}, V_{1y}, V_{1z}) = (0,0,0.5)$ and $(V_{2x}, V_{2y}, V_{2z})
= (0,0,-0.2)$ (in arbitrary units). We generated 100 events, and applied the
previous algorithm with $64^3$ cells in velocity space and a threshold of
$0.1$. We found that only two cells in velocity space are above threshold, and
they correspond exactly to the location of the simulated streams.
Fig.~\ref{fig:pignose} plots the $(v_x,v_z)$ section of the reconstructed
velocity distribution. It is impressive that we were able to recover this
velocity distribution with only 100 events.

We leave further studies of our simple algorithm, and the development of other
algorithms, to future work.

\section{Conclusions}

Directional detectors for WIMP dark matter searches will be able to measure not
only the energy but also the direction of the nuclear recoils caused by the
elastic scattering of galactic WIMPs with nuclei in the detector. This
directional capability will help in separating a WIMP signal from background,
and will also provide a measurement of the recoil momentum spectrum as compared
to just the recoil energy spectrum.

To gain insight into the properties of recoil momentum spectra, we have devised
a simple formalism for the analytic computation of recoil momentum spectra from
WIMP velocity distributions. Mathematically, the recoil momentum spectrum is
the 3-dimensional Radon transform of the velocity distribution.

Well-established mathematical properties of the Radon transform allow the
computation of analytical expressions for recoil spectra associated to several
common WIMP velocity distributions. As examples we presented recoil
spectra for a WIMP stream, a Maxwellian, a truncated Maxwellian, and a
non-isotropic Gaussian. We found in particular that a stream of WIMPs produces
a characteristic spherical pattern of nuclear recoils. A Maxwellian
distribution gives instead a smooth recoil pattern. Other velocity
distributions lead to more complicated spectra.

The analytic expressions we found for the nuclear recoil spectra will
facilitate the discrimination of different velocity distributions through
likelihood analysis.  In addition, it may be possible to invert the measured
momentum spectrum to reconstruct the local WIMP velocity distribution from
data. For this purpose, we have presented an algorithm to recover the velocity
distribution from a small number of recorded events. We have successfully
recovered a simulated velocity distribution with just 100 generated events.

We expect that the tools we have presented will be useful for the design and
analysis of directional WIMP detectors.

\acknowledgments

This research was supported in part by the National Science Foundation under
Grant No.\ PHY99-07949 at the Kavli Institute for Theoretical Physics,
University of California, Santa Barbara.

\appendix
\section{Some mathematics of the Radon transform}

In this appendix we collect some useful mathematical properties of the
3-dimensional Radon transform.  We denote the 3-dimensional Radon transform of
a function $f(\myvec{v})$ by $\radon{f}(w,\uvec{w}) $. It is defined by
\begin{equation}
  \label{eq:def}
  \radon{f}(w,\uvec{w}) = \int \delta( w - \uvec{w} \cdot \myvec{v} ) 
  f(\myvec{v}) d^3 v .
\end{equation}
It is easy to see that the Radon transform is linear,
\begin{equation}
  \radon{f_1+f_2} = \radon{f_1} + \radon{f_2}.
\end{equation}

\subsection{A remark on notation}
\label{sec:notation}

One may be tempted to write $\radon{f}(\myvec{w})$ for $\radon{f}(w,\uvec{w})$,
after all $\myvec{w} = w \uvec{w}$. This notation may however be ambiguous and
should be used with care. Indeed, one must keep in mind that the Radon
transform as defined in Eq.~(\ref{eq:def}) is a function of the magnitude $w$
and the direction $\uvec{w}$ separately. In other words, one may have
$\radon{f}(0,\uvec{w}) \ne \radon{f}(0,\uvec{w}')$ for $\uvec{w} \ne
\uvec{w}'$. Namely, $\radon{f}(0,\uvec{w})$ may assume different values for
different directions. This will not be reflected in the notation
$\radon{f}(\myvec{w})$. The latter would read $\radon{f}(0)$ at the origin,
independently of the direction $\uvec{w}$. In other words,
$\radon{f}(\myvec{w})$ would be a multiple-valued function at the origin.
Mathematically, the distinction between $\radon{f}(w,\uvec{w})$ and
$\radon{f}(\myvec{w})$ is important, and is expressed by saying that
$\radon{f}(w,\uvec{w})$ is defined on $ \mathbb{R} \times S^2$ while
$\radon{f}(\myvec{w})$ is defined on $\mathbb{R}^3$. For our application,
however, the distinction is of little concern, since the problematic origin
$w=0$ corresponds to the region of vanishingly small recoil momenta, which is
experimentally inaccessible. We have nevertheless used the mathematically
correct notation throughout for clarity.

\subsection{Change of coordinates}

Under linear transformations of the coordinate axes, $\myvec{v}$ transforms as
\begin{equation}
  \myvec{v} \to \myvec{v}' = {\rm A} \myvec{v} + \myvec{b},
\end{equation}
where ${\rm A}$ is a $3\times 3$ non-singular matrix and $ \myvec{b}$ is a
constant vector. A velocity distribution function $f(\myvec{v})$ transforms so
as to keep the number of particles in a volume $d^3v$ invariant:
\begin{equation}
  f'(\myvec{v}') d^3 v' = f(\myvec{v}) d^3v .
\end{equation}
Hence,
\begin{equation}
  f'(\myvec{v}') = \frac{1}{ | \!\det {\rm A} |} 
   \, f\!\left({\rm A}^{-1} (\myvec{v}'-\myvec{b}) \right),
\end{equation}
where $\det {\rm A}$ is the determinant of ${\rm A}$ and ${\rm A}^{-1}$ is the
inverse of ${\rm A}$. To find the relation between the Radon transforms of
$f(\myvec{v})$ and $f'(\myvec{v}')$, we change integration variable from
$\myvec{v}'$ to $\myvec{v}$ in the definition, Eq.~(\ref{eq:def}),
\begin{eqnarray}
  \radon{f'}(w',\uvec{w}') & = & 
  \int \delta( w' - \uvec{w}' \cdot \myvec{v}' )  \, f'(\myvec{v}')  \, d^3 v'
  =
  \int \delta( w' - \uvec{w}' \cdot {\rm A} \myvec{v} - 
  \uvec{w}' \cdot \myvec{b}) 
  \, f(\myvec{v})  \, d^3 v
  \nonumber \\
  & = & 
  \int \delta( w' - \uvec{w}' \cdot \myvec{b} - 
  {\rm A}^{\!T} \uvec{w}' \cdot \myvec{v} )  \, 
  f(\myvec{v})  \, d^3 v =   
  \radon{f}(w'- \uvec{w}' \cdot \myvec{b},{\rm A}^{\!T} \uvec{w}') ,
\end{eqnarray}
where ${\rm A}^{\!T}$ is the transpose of ${\rm A}$.  Thus
\begin{equation}
  \radon{f'}(w',\uvec{w}') = 
  \radon{f}(w'- \uvec{w}' \cdot \myvec{b}, \, {\rm A}^{\!T} \uvec{w}').
\end{equation}
In particular, under a pure rotation ${\rm R}$,
\begin{equation}
  \label{eq:rotate}
  \radon{f'}(w',\uvec{w}') =  \radon{f}(w', \, {\rm R}^{-1} \uvec{w}'),
\end{equation}
and under a pure translation $\myvec{b}$,
\begin{equation}
  \label{eq:transl}
  \radon{f'}(w',\uvec{w}') = \radon{f}(w'- \uvec{w}' \cdot \uvec{b}, 
  \, \myvec{w}').
\end{equation}

\subsection{Transformation of derivatives}

The following relations hold for derivatives of the Radon transform (here
$\myvec{v}=(v_1,v_2,v_3)$ and $\myvec{w}=(w_1,w_2,w_3)$) 
\begin{eqnarray}
  &&  \radon{\frac{\partial f}{\partial v_k}} = 
  \frac{w_k}{w} \frac{\partial\radon{f}}{\partial w} , \\
  &&  \frac{\partial\radon{f}}{\partial w_k} = - \frac{1}{w}
  \frac{\partial}{\partial w}  \radon{v_k f}.
\end{eqnarray}

\subsection{Integration over angles}

We find the following expression for the integral of the Radon transform $
\radon{f}(w,\uvec{w}) $ over the directions $ \uvec{w} $:
\begin{eqnarray}
  \label{eq:etaave}
  \int \radon{f}(w,\uvec{w}) d\Omega_w &=&
  \int \int \delta ( \myvec{v} \cdot \uvec{w} - w) 
  \, f(\myvec{v}) \, d^3 v \, d\Omega_w =
  \int \left[ \int \delta ( \myvec{v} \cdot \uvec{w} - w) \, d\Omega_w \right]
  \, f(\myvec{v}) \, d^3 v \nonumber\\ &=&
  \int \left[ 2 \pi \int_{-1}^{1}
    \delta ( v \cos\gamma - w) \, d\!\cos\gamma \right]
  \, f(\myvec{v}) \, d^3 v =
  \int \frac{2 \pi}{v} \, \theta(v-|w|) \,
  \, f(\myvec{v}) \, d^3 v \nonumber\\ &=&
  2 \pi  \int_{v>|w|} \frac{f(\myvec{v})}{v} \,\,  d^3v .
\end{eqnarray}

\subsection{Fourier slice theorem}

There is a connection between the Radon transform and the Fourier transform.
Taking the Fourier transform of the definition, Eq.~(\ref{eq:def}), with
respect to $w$ at fixed $\uvec{w}$ gives
\begin{equation}
  \label{eq:fst}
  \int_{-\infty}^{+\infty} d\lambda e^{i\lambda t} 
  \radon{f}(\lambda,\uvec{w}) =
  \int f(\myvec{v}) e^{i t \uvec{w} \cdot \myvec{v} } d^3v .
\end{equation}
This equation goes under the name of Fourier slice theorem.  The right hand
side is just the Fourier transform of $f(\myvec{v})$ evaluated at $t \uvec{w}$,
while the left hand side is the Fourier transform of $ \radon{f}(w,\uvec{w}) $
at fixed $ \uvec{w} $.

Inverting the Fourier transform in the left hand side of the Fourier slice
theorem, we have 
\begin{equation}
  \label{eq:fst2}
  \radon{f}(w,\uvec{w}) = \frac{1}{2 \pi} \int_{-\infty}^{\infty} dt 
  e^{-i w t} \int f(\myvec{v}) e^{i t \uvec{w} \cdot \myvec{v} } d^3v .
\end{equation}
This alternative expression of the Radon transform actually serves as its
definition when functions are replaced by distributions (in the mathematical
sense, see Ref.~\cite{rk96}).

\subsection{Expansion into spherical harmonics}

Let us expand $f(\myvec{v})$ and its Radon transform $\radon{f}(w,\uvec{w})$
into spherical harmonics $Y_{lm}(\uvec{v})$ and $Y_{lm}(\uvec{w})$,
respectively. We have
\begin{equation}
\label{eq:sph1}
f(\myvec{v}) = \sum_{lm} f_{lm}(v) Y_{lm}(\uvec{v}) , 
\end{equation}
and
\begin{equation}
\label{eq:sph2}
\radon{f}(w,\uvec{w}) = \sum_{lm} \radon{f}_{lm}(w) Y_{lm}(\uvec{w}) .
\end{equation}
The coefficients of the spherical harmonic expansions are related by
\begin{equation}
  \label{eq:etalm}
  \radon{f}_{lm}(w) = 2 \pi \int_{w}^{\infty} P_l\!\left(\frac{w}{v}\right) 
  f_{lm}(v) v dv .
\end{equation}
where $P_l(x)$ is a Legendre polynomial of order $l$.  These expressions are
useful when the velocity distributions are not isotropic.

Eq.~(\ref{eq:etalm}) can be proven using the decomposition of the
$\delta$-function in Legendre polynomials
\begin{equation}
  \delta( \uvec{v} \cdot \uvec{w} - t) = \theta(1-t) 
  \sum_l \frac{2l+1}{2}
  P_l(t) \, P_l(\uvec{v} \cdot \uvec{w}) ,
\end{equation}
the addition theorem for spherical harmonics
\begin{equation}
  P_l(\uvec{v} \cdot \uvec{w}) = \frac{4\pi}{2l+1} \sum_{m=-l}^{l}
  Y^*_{lm}(\uvec{v}) Y_{lm}(\uvec{w}) ,
\end{equation}
and the orthogonality of the spherical harmonics
\begin{equation}
  \int Y^*_{l'm'}(\uvec{v}) Y_{lm}(\uvec{v}) d\Omega_v = \delta_{l'l}
  \delta_{m'm} ,
\end{equation}
which lead to the relation
\begin{equation}
  \int \delta( \uvec{v} \cdot \uvec{w} - t) Y_{lm}(\uvec{v}) 
  d\Omega_v = 2 \pi \theta(1-t)
  P_l(t) Y_{lm}(\uvec{w}) .
\end{equation}

\section{Inversion formulas for the Radon transform}

\subsection{Inversion using the Laplacian}

An inversion formula for the Radon transform is
\begin{equation}
  \label{eq:inv1}
  f(\myvec{v}) = - \frac{1}{8\pi^2} \frac{\partial^2}{\partial \myvec{v}^2}
  \int \radon{f}(\myvec{v} \cdot \uvec{w}, \uvec{w}) d\Omega_w ,
\end{equation}
where $\partial^2/\partial \myvec{v}^2$ is the Laplacian in $\myvec{v}$. 
It can also be written in terms of $\radon{f}''(w,\uvec{w}) = 
\partial^2\radon{f}(w,\uvec{w}) / \partial w^2$ as
\begin{equation}
  \label{eq:inv1b}
  f(\myvec{v}) = - \frac{1}{8\pi^2} 
  \int \radon{f}''(\myvec{v} \cdot \uvec{w}, \uvec{w}) d\Omega_w ,
\end{equation}
The inversion formula (\ref{eq:inv1}) can be proven by inverting the Fourier
transform of $f(\myvec{v})$ in the Fourier slice theorem, Eq.~(\ref{eq:fst}),
then integrating separately in $t$ and $\uvec{w}$, and finally using the
relation
\begin{equation}
  \int_0^\infty t^2 e^{itx} dt = - \pi \delta^{(2)}(x) ,
\end{equation}
where $\delta^{(2)}(x)$ is the second derivative of the Dirac
$\delta$-function.

\subsection{Inversion through spherical harmonics}

Another inversion method is through an expansion in spherical harmonics. 
Referring to Eqs.~(\ref{eq:sph1}) and~(\ref{eq:sph2}), one can prove the
following inversion formula
\begin{equation}
  \label{eq:flm}
  f_{lm}(v) = -\frac{1}{2 \pi v} \int_{0}^{v} P_l\!\left(\frac{w}{v}\right)
  \radon{f}_{lm}''(w) dw ,
\end{equation}
where $P_l(x)$ is a Legendre polynomial and $\radon{f}_{lm}''(w) =
d^2\radon{f}_{lm}/dw^2$, the second derivative of $\radon{f}(w)$ with respect
to the modulus of $w$.

Eq.~(\ref{eq:flm}) is proven along the same lines as Eq.~(\ref{eq:etalm}),
starting from Eq.~(\ref{eq:inv1b}) written as
\begin{equation}
  f(\myvec{v}) = - \frac{1}{8\pi^2} 
  \int \delta( w - \myvec{v} \cdot \uvec{w}) \radon{f}''(w, \uvec{w}) dw 
  d\Omega_w .
\end{equation}

\subsection{Inversion through Fourier transforms}

The Fourier slice theorem, Eq.~(\ref{eq:fst}), can be made into an algorithm
for the numerical evaluation of the inverse Radon transform. Typically one
would use fast Fourier transforms. 

\subsection{Inversion through the adjoint operator}

Eq.~(\ref{eq:inv1}) can also be made into an algorithm. For each given
$\myvec{v}$, the integration in $d\Omega_w$ amounts to an integration over the
sphere of diameter $\myvec{v}$ and passing through the origin (a ``stream
sphere''), with integration measure $d\cos(\theta/2)d\phi$ in spherical
coordinates centered at the center of the sphere and north pole in $\myvec{v}$.
The final Laplacian can be computed numerically as the difference between the
central value and the average value of its six nearest neighbors.

\subsection{Algebraic inversion via discretization}

An algebraic inversion method is the following \cite[p.\ 96]{rk96}.  Suppose
that the values $\radon{f}_j$, $j=1,\dots,N$, corresponding to the points
$\myvec{w}_j$ are known. In medical applications, the points $\myvec{w}_j$ form
a grid or other structure in space, and the $\radon{f}_j$'s are the measured
signal intensities. In our case, the number of detected events may be quite
small, in which case we may let $\myvec{w}_j$ be the actual measurement of a
nuclear recoil momentum, with $j$ varying over the number of events, and
$\radon{f}_j = 1/\varepsilon_j$, where $\varepsilon_j$ is the efficiency for
detecting event $j$.

By definition of Radon transform we have
\begin{equation}
  \label{eq:etaj}
  \int_{P_j} f(\myvec{v}) d^3 v = \radon{f}_j ,
\end{equation}
where the integral is taken over the plane in $\myvec{v}$-space defined by the
equation
\begin{equation}
  P_j : \quad \myvec{v} \cdot \uvec{w} = w .
\end{equation}
$P_j$ is the plane orthogonal to the recoil direction $\uvec{w}$ and at a
distance $w$ from the origin. Now suppose that $f(\myvec{v})$ has compact
support, meaning that it vanishes for $|\myvec{v}|>$ something. This is a
technical simplification that is valid in practice since real velocity
distributions are always truncated at some large velocity (e.g.\ at the escape
speed from the galaxy). Divide the $\myvec{v}$-space into small cells $S_m$,
$m=1,\dots,M$, and assume that $f(\myvec{v})$ is constant over each of these
small cells, with value $f_m$ in cell $S_m$. This is the discretizing
approximation. Let
\begin{equation}
  a_{jm} = {\rm area}( S_m \cap P_j ) ,
\end{equation}
be the area of the intersection of the plane $P_j$ with the cell $S_m$ (see
Appendix C for an explicit expression).  A discretized version of
Eq.~(\ref{eq:etaj}) is then
\begin{equation}
  \sum_m a_{jm} f_m = \radon{f}_j .
\end{equation}
In matrix form
\begin{equation}
  A f = \radon{f} ,
\end{equation}
where $A=(a_{jm})$ is an $N\times M$ matrix, $f=(f_1,\dots,f_M)^T$ and
$\radon{f}=(\radon{f}_1,\dots,\radon{f}_N)^T$. 
This is a system of linear equations for $f$ that can be solved by
inverting $A$.  Since few $a_{jm}$ differ from zero, $A$ is a sparse matrix,
and it is convenient to solve this system iteratively. Fix $\omega$,
$0<\omega<2$. Let the initial guess be $f^{(0)}$ and the $k$-th update be
$f^{(k)}$. From $f^{(k)}$ compute the following vectors successively
\begin{eqnarray}
  f^{(k)}_0 & = & f^{(k)}, \\
  f^{(k)}_j & = & f^{(k)}_{j-1} + \frac{\omega}{a_j^2} 
  \left( \radon{f}_j - a_j^T
    f^{(k)}_{j-1} \right) a_j, \quad j=1,\dots,N .
\end{eqnarray}
Here $a_j^2 = \sum_m a_{jm}^2$ and $a_j = (a_{j1}, \dots, a_{jM})$. Finally let
the next update be $f^{(k+1)} = f^{(k)}_N$. Ref.~\cite{rk96} attributes this
method to Kaczmarz. 

\section{Area of the intersection between a plane and a cell}

For future reference, we give here the expression for the area of the
intersection of a plane with a rectangular cell. 

Assume the $(x,y,z)$ space is divided into rectangular cells of sides $h_x$,
$h_y$, and $h_z$ along $x$, $y$, and $z$, respectively. Let the $(i,j,k)$-th
cell $S_{ijk}$ be centered in $(x_0+i h_x, y_0+j h_y, z_0+k h_z)$,
\begin{equation}
  S_{ijk} : \quad  
  \left\{ 
    \begin{array}{l}
      \left(i-\frac{1}{2}\right) h_x + x_0 < x < 
      \left(i+\frac{1}{2}\right) h_x + x_0 ,  \vspace{5pt} \\ 
      \left(j-\frac{1}{2}\right) h_y + y_0 < y < 
      \left(j+\frac{1}{2}\right) h_y + y_0 ,  \vspace{5pt} \\
      \left(k-\frac{1}{2}\right) h_z + z_0 < z < 
      \left(k+\frac{1}{2}\right) h_z + z_0 .
    \end{array}
  \right.
\end{equation}
Let $P$ be the plane defined by
\begin{equation}
  P : \quad \omega_x x + \omega_y y + \omega_z z = p .
\end{equation}
Then the area of the intersection of the plane $P$ with the $(i,j,k)$-th cell
$S_{ijk}$ is
\begin{equation}
  {\rm area}( S_{ijk} \cap P) \, = \, h_x h_y h_z \times 
  \left\{
    \begin{array}{ll}
      \displaystyle
      0, & \quad\hbox{if $P\sqrt{ K_1^2 + K_2 ^2 + K_3 ^2 } > \sqrt{3}/2$,}  
      \vspace{5pt} \\
      \displaystyle
      \frac{1}{K_3}, & \quad\hbox{if $P \le |K|$ and $K\ge 0$,}  
      \vspace{5pt} \\
      \displaystyle
      \frac{1}{K_3}- \frac{P^2+K^2}{K_1K_2K_3}, & 
      \quad\hbox{if $P \le |K|$ and $K<0$,}
      \vspace{5pt} \\
      \displaystyle
      \frac{1}{K_3}- \frac{(K-P)^2}{2K_1K_2K_3}, & 
      \quad\hbox{if $P > |K|$,}
    \end{array}
  \right.
  \label{eq:explicitarea}
\end{equation}
where
\begin{eqnarray}
  P & = & \bigl| p - \omega_x (x_0+i h_x) - \omega_y (y_0+j h_y) - \omega_z
  (z_0+k h_z) \bigr| , \\
  K & = & \frac{1}{2} \left( K_3 - K_2 - K_1 \right) , 
\end{eqnarray}
and $K_1$, $K_2$, and $K_3$ are the quantities $| h_x \omega_x |$, $| h_y
\omega_y |$, and $ | h_z \omega_z |$ sorted in order of increasing magnitude, $
K_1 \le K_2 \le K_3$. The last case in Eq.~(\ref{eq:explicitarea}) becomes 
\begin{equation}
  \frac{ K_2+K_3-2P }{ 2 K_2 K_3 }
\end{equation}
in the limit of small $K_1$.

\newpage

\newpage

\newpage

\newpage

\end{document}